\documentclass[twocolumn]{aastex61}
\usepackage{amsmath}

\begin{document}

\title{A GRB afterglow model consistent with hypernova observations}

\author{R.~Ruffini}

\affiliation{ICRA and Dipartimento di Fisica, Sapienza Universit\`a di Roma, P.le Aldo Moro 5, 00185 Rome, Italy.}
\affiliation{ICRANet, P.zza della Repubblica 10, 65122 Pescara, Italy.}
\affiliation{Universit\'e de Nice Sophia Antipolis, CEDEX 2, Grand Ch\^{a}teau Parc Valrose, Nice, France.}
\affiliation{ICRANet-Rio, Centro Brasileiro de Pesquisas F\'isicas, Rua Dr. Xavier Sigaud 150, 22290--180 Rio de Janeiro, Brazil.}

\author{M.~Karlica}
\affiliation{ICRA and Dipartimento di Fisica, Sapienza Universit\`a di Roma, P.le Aldo Moro 5, 00185 Rome, Italy.}
\affiliation{ICRANet, P.zza della Repubblica 10, 65122 Pescara, Italy.}
\affiliation{Universit\'e de Nice Sophia Antipolis, CEDEX 2, Grand Ch\^{a}teau Parc Valrose, Nice, France.}

\author{N.~Sahakyan}
\affiliation{ICRANet, P.zza della Repubblica 10, 65122 Pescara, Italy.}
\affiliation{ICRANet-Armenia, Marshall Baghramian Avenue 24a, Yerevan 0019, Armenia.}

\author{J.~A.~Rueda}
\affiliation{ICRA and Dipartimento di Fisica, Sapienza Universit\`a di Roma, P.le Aldo Moro 5, 00185 Rome, Italy.}
\affiliation{ICRANet, P.zza della Repubblica 10, 65122 Pescara, Italy.}
\affiliation{ICRANet-Rio, Centro Brasileiro de Pesquisas F\'isicas, Rua Dr. Xavier Sigaud 150, 22290--180 Rio de Janeiro, Brazil.}

\author{Y.~Wang}
\affiliation{ICRA and Dipartimento di Fisica, Sapienza Universit\`a di Roma, P.le Aldo Moro 5, 00185 Rome, Italy.}
\affiliation{ICRANet, P.zza della Repubblica 10, 65122 Pescara, Italy.}

\author{G.~J.~Mathews}
\affiliation{ICRANet, P.zza della Repubblica 10, 65122 Pescara, Italy.}
\affiliation{Center for Astrophysics, Department of Physics, University of Notre Dame, Notre Dame, IN, 46556, USA.}

\author{C.~L.~Bianco}
\affiliation{ICRA and Dipartimento di Fisica, Sapienza Universit\`a di Roma, P.le Aldo Moro 5, 00185 Rome, Italy.}
\affiliation{ICRANet, P.zza della Repubblica 10, 65122 Pescara, Italy.}

\author{M.~Muccino}
\affiliation{ICRA and Dipartimento di Fisica, Sapienza Universit\`a di Roma, P.le Aldo Moro 5, 00185 Rome, Italy.}
\affiliation{ICRANet, P.zza della Repubblica 10, 65122 Pescara, Italy.}

\begin{abstract}
We describe the afterglows of the long gamma-ray-burst (GRB) 130427A within the 
context of a binary-driven hypernova (BdHN). The afterglows originate from the interaction between a newly born neutron star ($\nu$NS), created by an Ic supernova (SN), and a mildly 
relativistic ejecta of a hypernova (HN). Such a HN in turn results from 
the impact of the GRB on the original SN Ic. The mildly 
relativistic expansion velocity of the afterglow ($\Gamma \sim 3$) is determined, using our model independent approach, from the thermal 
emission between $196$~s and $461$~s. The power-law 
in the optical and X-ray bands of the afterglow is shown to arise from the synchrotron 
emission of relativistic electrons in the expanding magnetized HN 
ejecta. Two components contribute to the injected energy: the kinetic 
energy of the mildly relativistic expanding HN and the rotational 
energy of the fast rotating highly magnetized $\nu$NS. 
We reproduce the afterglow in all wavelengths from 
the optical ($10^{14}$~Hz) to the X-ray band ($10^{19}$~Hz) over times 
from $604$~s to $5.18\times 10^6$~s relative to the Fermi-GBM trigger. 
Initially, the emission is dominated by the loss of kinetic energy of 
the HN component.  After $10^5$~s the emission is dominated by the loss 
of rotational energy of the $\nu$NS, for which we adopt an initial 
rotation period of $2$~ms and a dipole {plus} quadrupole magnetic field of 
$\lesssim \! 7\times 10^{12}$~G {or} $\sim \! 10^{14}$~G. 
This scenario {with a progenitor composed of a CO$_{\rm core}$ and a NS companion} differs from the traditional ultra-relativistic-{jetted} treatments of the afterglows {originating from a single black hole.}
\end{abstract}


%%%%%%%%%%%%%%%%%%%%%%%%%%%%%%%%%%%%%%%%%%%%%%%%%%%%%%
%%%%%%%%%%%%%%%%%%%%%%%%%%%%%%%%%%%%%%%%%%%%%%%%%%%%%%
\section{Introduction}\label{sec:1}
%%%%%%%%%%%%%%%%%%%%%%%%%%%%%%%%%%%%%%%%%%%%%%%%%%%%%%
%%%%%%%%%%%%%%%%%%%%%%%%%%%%%%%%%%%%%%%%%%%%%%%%%%%%%%

It has been noted for almost two decades \citep{1998Natur.395..670G} that many long-duration GRBs show the presence of an associated unusually energetic supernova (SN) of type Ic (hypernova, HN) as well as of a long-lasting X-ray afterglow \citep{Costa1997}. Such HNe are unique in their spectral characteristics; they have no hydrogen and helium lines, suggesting that they are members of a binary system \citep{2009ARA&A..47...63S}. Moreover, these are broad-lined HNe suggesting the occurrence of energy injection beyond that of a normal type Ic SN \citep{2016MNRAS.457..328L}.

This has led to our suggestion \citep[e.g.][]{2001ApJ...555L.117R,2012A&A...548L...5I} of a model for long GRBs associated with SNe Ic. In this paradigm, the progenitor is a carbon-oxygen star (CO$_{\rm core}$) in a tight binary system with a neutron star (NS). As the CO$_{\rm core}$ explodes in a type Ic SN it produces a new NS (hereafter $\nu$NS) and ejects a remnant of { a} few solar masses, some of which is accreted onto the companion NS \citep{2012ApJ...758L...7R}. The accretion onto the companion NS is hypercritical, i.e. highly super-Eddington, reaching accretion rates of up to a tenth of solar mass per second, for the most compact binaries with orbital periods of { a} few minutes \citep{2014ApJ...793L..36F}. The NS gains mass rapidly, reaching the critical mass, within  { a} few seconds. The NS then collapses to a black hole (BH) with the consequent emission of the GRB \citep{2015PhRvL.115w1102F}. In this picture the BH formation and the associated GRB occurs some seconds {\it after} the initiation of the SN. The high temperature and density reached during the hypercritical accretion and the NS collapse lead to a copious emission of $\nu\bar\nu$ pairs which form an $e^+e^-$ pair plasma that drives the GRB \citep[see e.g.][]{2015ApJ...812..100B,2016ApJ...833..107B,2016ApJ...832..136R}. The expanding SN remnant is reheated and shocked by the injection of the $e^+e^-$ pair plasma from the GRB explosion \citep{2018ApJ...852...53R}. 

The shocked-heated SN, originally expanding at $0.2 c$, is transformed into an HN reaching expansion velocities up to {$0.94 c$} (see Sec.~\ref{sec:3}). A vast number of totally new physical processes are introduced that must be treated within a correct classical and quantum general relativistic approach \citep[see e.g.][and references therein]{2018ApJ...852...53R}. The ensemble of these processes, addressing causally disconnected phenomena, each characterized by specific world lines, ultimately leads to a specific Lorentz $\Gamma$ factor. This ensemble comprises the binary-driven hypernova (BdHN) paradigm \citep{2016ApJ...832..136R}.

In this article we extend this novel approach to the analysis of the BdHN afterglows. The existence of regularities in the X-ray luminosity of BdHNe, expressed in the observer cosmological rest-frame, has been previously noted leading to the Muccino-Pisani power-law behavior \citep{Pisani2013,2014A&A...565L..10R}. The aim of this article is to now explain the origin of these power-law relations and to understand their physical origin and their energy sources.

The kinetic energy of the mildly relativistic expanding HN at {$0.94 c$} following the $\gamma$-ray flares and the X-ray flares, as well as the overall plateau phase, appears to have a crucial role \citep{2014A&A...565L..10R}. Equally crucial appears to be the contribution of the rotational energy electromagnetically radiated by the $\nu$NS. As we show in this article, the power-law luminosity in the X-rays and in the optical wavelengths, expressed as a function of time in the GRB source rest-frame, could not be explained without their fundamental contribution. We here indeed assume that the afterglow originates from the synchrotron emission of relativistic electrons injected in the magnetized plasma of the HN, using both the kinetic energy of expansion and the electromagnetic energy powered by the rotational energy loss of the $\nu$NS~{(see Sec.~\ref{sec:4})}.

As an example, we apply this new approach to the afterglow of GRB 130427A associated to the SN 2013cq, in view of the excellent data available in X-rays, optical and radio wavelengths. We fit the spectral evolution of the GRB from 604 to $5.18 \times 10^6$~s and over the observed frequency bands from 10$^9$~Hz to 10$^{19}$~Hz. We present our simulations of the afterglow of GRB 130427A suggesting that a total energy { of order $\simeq 10^{53}$~erg has been injected into the electrons confined within the expanding magnetized HN.  This energy derives from  the kinetic energy of the HN and the rotational energy of the $\nu$NS with a rotation period 2~ms,  containing a dipole or quadrupole magnetic field of $(5$--$7)\times 10^{12}$ G or $10^{14}$~G.}

The article is organized as follows. In Sec.~\ref{sec:2} we summarize how the BdHN treatment compares and contrasts with the traditional collapsar-fireball model of the GRB afterglow which is based on a single ultra-relativistic jet. In Sec.~\ref{sec:3} we present the data reduction of GRB 130427A. In Sec.~\ref{sec:4} we examine the basic parameters of the $\nu$NS relevant for this analysis such as the rotation period, the mass, the rotational energy, and the magnetic field structure. We introduce in Sec.~\ref{sec:5} the main ingredients and equations relevant for the computation of the synchrotron emission of the relativistic electrons injected in the magnetized HN. In Sec.~\ref{sec:6} we set up the initial/boundary conditions to solve the model equations of Sec.~\ref{sec:5}. In Sec.~\ref{sec:7} we compare and contrast the results of the numerical solution of our synchrotron model, the theoretical spectrum and light-curve, with the afterglow data of GRB 130427A at early times $10^2~s \lesssim  t \lesssim 10^6$~s. We also show the role of the $\nu$NS in powering the late, $t \gtrsim 10^6$~s, X-ray afterglow. Finally, we { present} our conclusions in Sec.~\ref{sec:8} outlining some possible further observational predictions of our model.

%%%%%%%%%%%%%%%%%%%%%%%%%%%%%%%%%%%%%%%%%%%%%%%%%%%%%%
%%%%%%%%%%%%%%%%%%%%%%%%%%%%%%%%%%%%%%%%%%%%%%%%%%%%%%
\section{On BdHNe versus the traditional collapsar-fireball approach}\label{sec:2}
%%%%%%%%%%%%%%%%%%%%%%%%%%%%%%%%%%%%%%%%%%%%%%%%%%%%%%
%%%%%%%%%%%%%%%%%%%%%%%%%%%%%%%%%%%%%%%%%%%%%%%%%%%%%%

In \citet{2016ApJ...832..136R} it was established that there exist seven different GRB subclasses, all with binary systems as progenitors composed of various combinations of white dwarfs (WDs), CO$_{\rm cores}$, NSs and BHs, and that in only in three of these subclasses are BHs formed. Far from being just a morphological classification, the identification of these systems and their properties has been made possible by the unprecedented quality and { extent} of the data ranging from X-ray, to the $\gamma$-ray, to the GeV emission as well as in the optical and in the radio. A comparable effort has been progressing in the theoretical field by introducing new paradigms and developing consistently the theoretical framework.

The main {insight} gained from BdHN paradigm, one of the most numerous of the above seven subclasses, has been the successful identification, guided by the observational evidence, of a vast number of independent processes {of the GRB}. For each process the corresponding field equations have been integrated, obtaining their Lorentz $\Gamma$ factors as well as their space-time evolution. This is precisely what has been done in the recent publications for the {{ultrarelativistic prompt emission (UPE) in the first 10 seconds with Lorentz factor $\Gamma \sim 500$--$1000$, the hard X-ray flares (HXF) with $\Gamma \sim 10$ and for the mildly relativistic soft X-ray flares (SXF) with $\Gamma \sim 2-3$ \citep{2018ApJ...852...53R} with the extended thermal X-ray emission (ETE) { signaling}  the transformation of a SN into a HN \citep{transition}.}}

Here we extend the BdHN model to { the study of the afterglow.  As a prototype we utilize the data of GRB 130427A. { We point out for the first time:}
\begin{enumerate}
\item The role of the hypernova ejecta and of the rotation of the binary system in creating the condition for the occurrence of synchrotron emission, rooted in the pulsar magnetic field (see Sec.~\ref{sec:4}).
\item The fundamental role played by the pulsar like behavior of the $\nu$NS (see Fig.~\ref{fig:Lpulsar}) and its magnetic field to explain the fit of { a} synchrotron model based on the optical and X-ray data (see Fig.~\ref{fig:movinglimitsmodelb05e2min1e3max5e5}).
\item To develop a model of the afterglow consistent with the mildly relativistic expansion velocity measured in the afterglows following a model-independent procedure (see Eq.(\ref{betaNew}) and Fig.~\ref{pobb} in Sec.~\ref{sec:3}).
\end{enumerate}
}

{{In the current afterglow model \citep[see, e.g.,][and references therein]{1999PhR...314..575P,Meszaros2002,Meszaros2006,2015PhR...561....1K} it is tacitly assumed that a {\it single} ultra-relativistic regime extends all the way from the prompt emission, to the plateau phase, all the way to the GeV emission and to the latest power-law of the afterglow. This approach is clearly in contrast with the point 3 above.}}

%%%%%%%%%%%%%%%%%%%%%%%%%%%%%%%%%%%%%%%%%%%%%%%%%%%%%%
%%%%%%%%%%%%%%%%%%%%%%%%%%%%%%%%%%%%%%%%%%%%%%%%%%%%%%
\section{GRB 130427A data}\label{sec:3}
%%%%%%%%%%%%%%%%%%%%%%%%%%%%%%%%%%%%%%%%%%%%%%%%%%%%%%
%%%%%%%%%%%%%%%%%%%%%%%%%%%%%%%%%%%%%%%%%%%%%%%%%%%%%%

GRB 130427A is well-known for its high isotropic energy $E_{iso} \simeq 10^{54}$~erg, SN association and multi-wavelength observations \citep{2015ApJ...798...10R}. It triggered \textit{Fermi}-GBM at 07:47:06.42 UT on April 27 2013 \citep{2013GCN.14473....1V}, when it was within the field of view of \textit{Fermi}-LAT. A a long-lasting ($\sim 10^4$~s) burst of ultra-high energy ($100$ MeV--$100$ GeV) radiation was observed \citep{2014Sci...343...42A}. \textit{Swift} started to follow from 07:47:57.51 UT, $51.1$~s after the GBM trigger, observing a soft X-ray ($0.3$--$10$~keV) afterglow for more than $100$~days \citep{2014Sci...343...48M}. NuStar joined the observation during three epochs, approximately $\sim 1.2$, $4.8$ and $5.4$~days after the \textit{Fermi}-GBM trigger, providing rare hard X-ray ($3$--$79$~keV) afterglow observations \citep{2013ApJ...779L...1K}. Ultraviolet, optical, infrared, { and }radio observations were also performed by more than $40$ satellites and ground-based telescopes, within which \textit{Gemini-North}, NOT, \textit{William Herschel}, and VLT confirmed the redshift of $0.34$ \citep{2013GCN.14455....1L,2013GCN.14478....1X,2013GCN.14617....1W,2013GCN.14491....1F}, and NOT found the associated supernova SN 2013cq \citep{2013ApJ...776...98X}. We adopt the radio, optical and the GeV data from various published articles and GCNs \citep{2014ApJ...781...37P,2014Sci...343...48M,2013GCN.14473....1V,2013GCN.14475....1S,2013ApJ...776...98X,2015ApJ...798...10R}. The soft and hard X-rays, which are one { of} the main subjects of this paper, { were} analyzed from the original data downloaded from \textit{Swift} repository\footnote{\noindent \url{http://www.swift.ac.uk}} and \textit{NuStar} archive\footnote{\noindent \url{https://heasarc.gsfc.nasa.gov/docs/nustar/nustar_archive.html}}. We { followed} the standard data reduction procedure Heasoft 6.22 with relevant calibration files\footnote{\noindent \url{http://heasarc.gsfc.nasa.gov/lheasoft/}}, and {  the spectra were generated}  by XSPEC 12.9 \citep{Evans:2007iz,Evans:2009kx}. During the data reduction, the pile-up effect in the \textit{Swift}-XRT { were}  corrected for the first $5$ time bins (see Fig.~\ref{fig:Lx}) before $10^5$~s \citep{Romano:2006kt}. The NuStar spectrum at $388800$~s is inferred from the closest first $10000$~s of the \textit{NuStar} third epoch at $\sim 5.4$~days, by assuming { that} the spectra at these two times have the same cutoff power-law shape but different amplitudes. The amplitude at $388800$~s { was} computed by fitting the \textit{NuStar} light-curve. A K-correction { was} implemented for transferring observational data to the cosmological rest frame \citep{2001AJ....121.2879B}.

{The GRB afterglow emission in the BdHN model originates from a mildly relativistic expanding supernova ejecta. This has been confirmed by measuring the expansion velocity $\beta \sim 0.6-0.9$ (corresponding to the Lorentz gamma factor $\Gamma < 5$) within { the} early hunderds of seconds after the trigger from the observed thermal emission in the soft X-ray. For instance, \citet{2014A&A...565L..10R} finds a velocity of $\beta \sim 0.8$ for GRB 090618, and in \citep{2018ApJ...852...53R}, GRB 081008 is found { to have} a velocity $\beta \sim 0.9$. The optical signal at tens of days also implies a mildly relativistic velocity $\beta \sim 0.1$  \citep{1998Natur.395..670G,2006ARA&A..44..507W,2017AdAst2017E...5C}.} 

{The expanding velocity can be directly inferred from the observable X-ray thermal emission and is summarised from \citet{2018ApJ...852...53R}:}

\begin{multline}
\frac{\beta^5}{4 [ \ln (1+ \beta) - (1-\beta) \beta]^2} \left(\frac{1+\beta}{1-\beta}\right)^{1/2}= \\
\frac{D_L(z)}{1+z} \frac{1}{t_2-t_1} \left(\sqrt{\frac{F_\mathrm{bb,obs} (t_2)}{\sigma T_\mathrm{obs}^4(t_2)}} - \sqrt{\frac{F_\mathrm{bb,obs}(t_1)}{\sigma T_\mathrm{obs}^4(t_1)}}\right) ,
\label{betaNew}
\end{multline}

\begin{figure}
	\centering
	\includegraphics[width=\hsize]{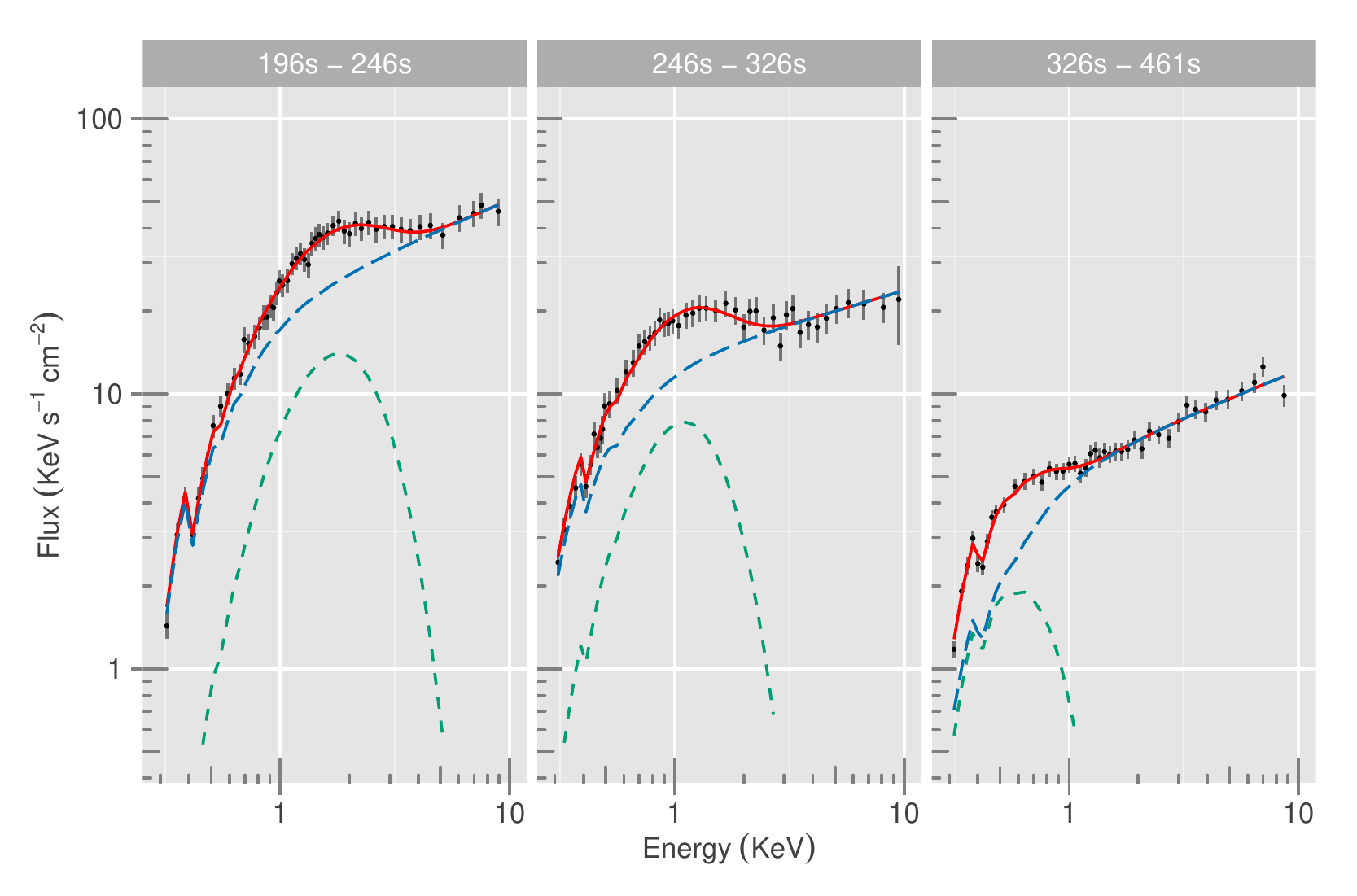}
	\caption{{ Spectral fitting  \citep{2015ApJ...798...10R}} of three time intervals (196s - 246s, 246s - 326s, 326s - 461s) in the \textit{Swift}-XRT band (0.3 keV - 10 keV). Black points presents the spectral data with H absorption, green dashed line is the fitted thermal component, blue long-dashed line is the power-law component, and red line is the sum of two components. Clearly the temperature and the thermal flux drop along the time.}
	\label{pobb}
\end{figure}

{The left term is a function of velocity $\beta$, the right term is from observables, $D_L(z)$ is the luminosity distance for redshift $z$. From the observed thermal flux $F_\mathrm{bb,obs}$ and temperature  $T_\mathrm{obs}$ at time $t_1$ and $t_2$, the velocity $\beta$ can be inferred. This model independent equation valid in Newtonian and relativistic regimes is general. The results inferred do not agree with the ones of the fireball model \citep{2002MNRAS.336.1271D,2007ApJ...664L...1P}, coming from a ultra-relativistic shockwave.}

{ Indeed,  GRB 130427A is a well-known example of a GRB associated with SN \citep{2013ApJ...776...98X}.  For this GRB an X-ray thermal emission has been  found between $196$--$461$~s \citep{2015ApJ...798...10R}.   The spectral evolution of this source is presented in Figure \ref{pobb}. From the  best fit, we obtain a temperature in the observer's frame that drops in time from $0.46$~keV to $0.13$~keV. The thermal flux also diminishes in time. }

{  From  Eq.~ (\ref{betaNew}), we obtain} a radius in the laboratory frame that increases from $1.67^{+0.43}_{-0.28} \times 10^{13}$~cm to $1.12^{+0.49}_{-0.33} \times 10^{14}$~cm. The velocity inferred from the first and second spectra is $\beta = 0.85^{+0.06}_{-0.10}$, from the second and third spectra increases to $\beta = 0.96^{+0.02}_{-0.03}$. The average velocity of the entire duration of thermal emission is $\beta = 0.94^{+0.03}_{-0.05}$, corresponding to a Lorentz factor $\Gamma = 2.98^{+1.20}_{-0.79}$, at an average radius $3.50^{+1.46}_{-0.97} \times 10^{13}$~cm. At later observer's time around $16.7$ days after the GRB trigger, the mildly relativistic velocity $\sim 32,000 ~ \mathrm{km \, s^{-1}}$ ($\beta \sim 0.1$) of the afterglow is measured from the line of Fe II 5169 \citep{2013ApJ...776...98X}. { Both the mildly relativistic velocities and the small radii are inferred directly from the observations and agree with the required properties of the} BdHN model.

The above data are in contrast with the traditional fireball model [e.g.~\citep{1999PhR...314..575P},] which involves a shockwave with a high Lorentz factor $\Gamma \sim 500$ continuously expanding and generating the prompt emission at { a radius of} $\sim 10^{15}$~cm, and then the afterglow at { a lab-frame  radius of} $>10^{16}$~cm. Therefore, any model of the afterglow with ultra relativistic velocity following after { the }UPE does not conform to the stringent observational constraints.

{ One is left, therefore,  with the task of developing a consistent afterglow model with a mildly relativistic expansion that is compatible with this clear observational evidence that the afterglow arises from mildly relativistic ejecta. That is the purpose of the present work}

%%%%%%%%%%%%%%%%%%%%%%%%%%%%%%%%%%%%%%%%%%%%%%%%%%%%%
%%%%%%%%%%%%%%%%%%%%%%%%%%%%%%%%%%%%%%%%%%%%%%%%%%%%%
\section{Role of the new fast-rotating NS in the energetics and properties of the GRB afterglow}\label{sec:4}
%%%%%%%%%%%%%%%%%%%%%%%%%%%%%%%%%%%%%%%%%%%%%%%%%%%%%
%%%%%%%%%%%%%%%%%%%%%%%%%%%%%%%%%%%%%%%%%%%%%%%%%%%%%

{ Angular momentum conservation} implies that the $\nu$NS should be rapidly rotating. For { example},  the gravitational collapse of an iron core of radius $R_{\rm Fe}\sim 5\times 10^8$~cm of a carbon-oxygen { progenitor} star leading to a SN Ic, rotating with { an initial  period  of $P\sim 5$~min, implies a rotation period $P = (R_{\rm NS}/R_{\rm Fe})^2 P_{\rm CO} \sim 1$~ms for the newly formed neutron star. Thus, one  expects the $\nu$NS to have } a large amount of rotational energy available to power the SN remnant. In order to evaluate { such a} rotational energy we need to know the {  structure  of fast rotating NSs. This} we adopt from \citet{2015PhRvD..92b3007C}. 

The structure of NSs in uniform rotation is obtained by numerical integration of the Einstein equations in axial symmetry and the stability sequences are described by two parameters, e.g.: the baryonic mass (or the gravitational mass/central density) and the angular momentum (or the angular velocity/polar to equatorial radius ratio).
The stability of the star is bounded by (at least) two limiting conditions \citep[see e.g.][for a review]{2003LRR.....6....3S}. The first is the mass-shedding or Keplerian limit: for a given mass (or central density) there is a configuration whose angular velocity equals { that} of a test particle in circular orbit at the stellar equator. Thus, the matter at the stellar surface is marginally { bound { so that} any small perturbation causes mass loss bringing the star back to stability or to a point of dynamical instability}. The second is the secular axisymmetric instability: in this limit the star becomes unstable against axially symmetric perturbations and is expected to evolve first quasi-stationarily { toward}  a dynamical instability point where gravitational collapse { ensues}. This instability sequence thus leads to the NS critical mass and it can be obtained via the turning-point method by \citet{1988ApJ...325..722F}. 

{ In} \citet{2015PhRvD..92b3007C} the values of the critical mass were obtained for the NL3, GM1 and TM1 { equations of state (EOS)} and the following fitting formula was found to describe them with a maximum error of 0.45\%:
\begin{equation}\label{eq:Mcrit}
M_{\rm NS}^{\rm crit}=M_{\rm crit}^{J=0}(1 + C j_{\rm NS}^a),
\end{equation}
where $j_{\rm NS}\equiv c J_{\rm NS}/(G M_\odot^2)$ is a dimensionless angular momentum parameter, $J_{\rm NS}$ is the NS angular momentum, $C$ and $a$ are parameters that depend on the nuclear EOS, and $M_{\rm crit}^{J=0}$ is the critical mass in the non-rotating case (see Table \ref{tb:StaticRotatingNS}).
\begin{table*}
\centering
\caption{Critical mass (and corresponding radius) obtained in \citet{2015PhRvD..92b3007C} for selected parameterizations of the nuclear EOS.}\label{tb:StaticRotatingNS}
\begin{tabular}{cccccccc}
\hline \hline
EOS  &  $M_{\rm crit}^{J=0}$~$(M_{\odot})$ & $R_{\rm crit}^{J=0}$~(km) & $M_{\rm max}^{J\neq 0}$~$(M_{\odot})$ & $R_{\rm max}^{J\neq 0}$~(km) &$a$&$C$ & $P_{\rm min}$~(ms) \\
\hline
NL3 & $2.81$&$13.49$ &$3.38$ & 17.35 & $1.68$&$0.006$ & $0.75$\\
GM1 & $2.39$&$12.56$ &$2.84$& 16.12&$1.69$&$0.011$ & $0.67$\\
TM1 & $2.20$ &$12.07$ &$2.62$ & 15.98 &$1.61$&$0.017$ & $0.71$\\  
\hline
\end{tabular}
\tablecomments{In the last column we list the rotation period of the fastest possible configuration which corresponds to that of the critical mass configuration (i.e. secularly unstable) that intersects the Keplerian mass-shedding sequence.}
\end{table*}

\begin{figure}
\centering
\includegraphics[width=\hsize,clip]{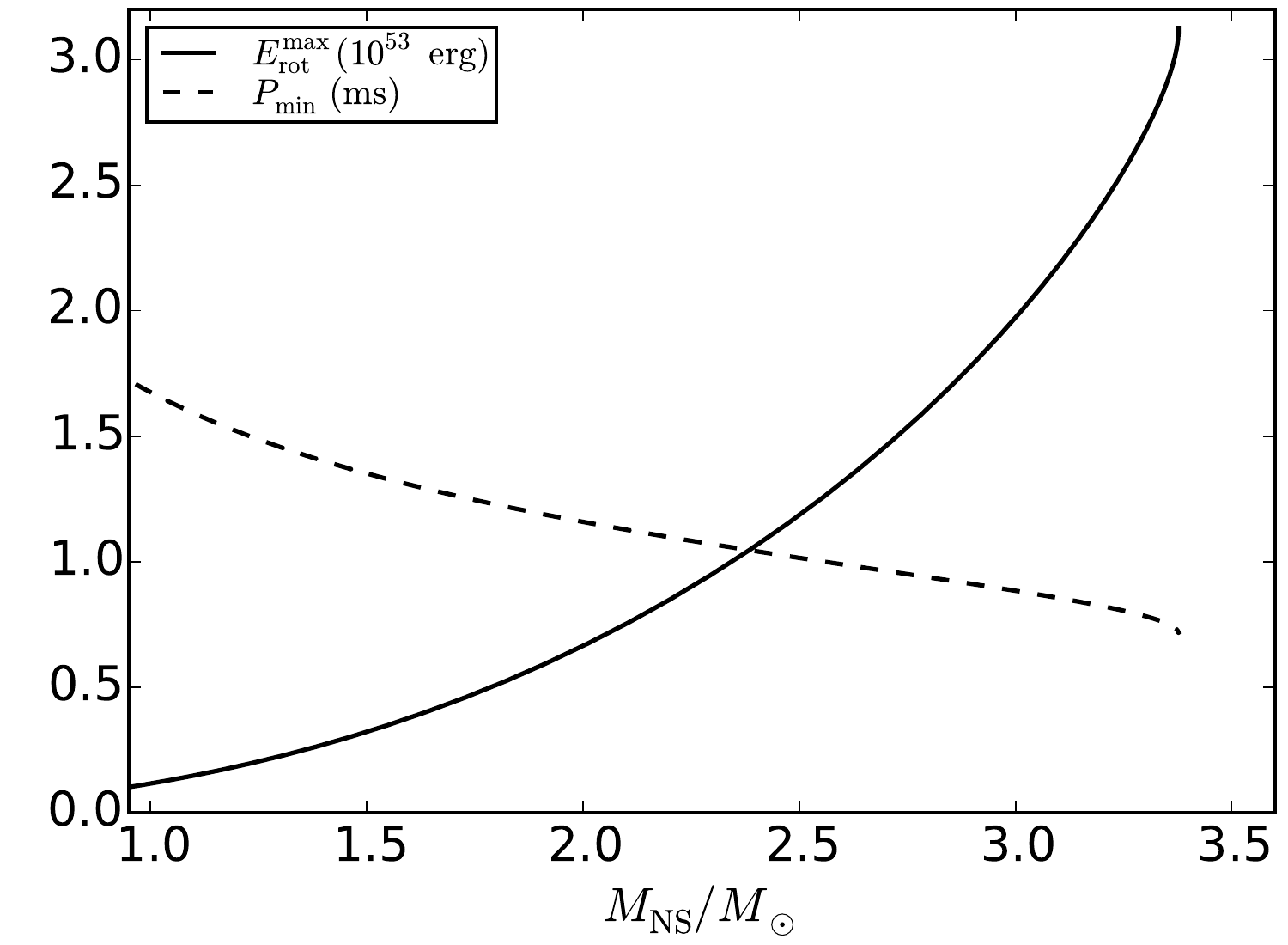}
\caption{Rotational energy and period of NSs along the Keplerian sequence for the NL3 EOS.}\label{fig:ErotvsM}
\end{figure}

The configurations lying along the Keplerian sequence are also the maximally rotating ones (given a mass or central density). The fastest rotating NS is the configuration at the crossing point between the Keplerian and the secular axisymmetric instability sequences. Fig.~\ref{fig:ErotvsM} shows the minimum rotation period and the rotational energy as a function of the NS gravitational mass for the NL3 EOS.

We turn now to the magnetosphere properties. Within the traditional model of pulsars \citep{1969ApJ...157..869G}, in a rotating, highly magnetized NS, a corotating magnetosphere is enforced up to a maximum distance $R_{\rm lc}=c/\Omega=c P/(2\pi)$, where $c$ is the speed of light and $\Omega$ is the angular velocity of the star. This defines the so-called light cylinder since corotation at larger distances { implies} superluminal velocities of the magnetospheric particles. The last $B$-field line closing within the corotating magnetosphere is located at an angle $\theta_{\rm pc} = \arcsin(\sqrt{R_{\rm NS}/R_{\rm lc}})\approx \sqrt{R_{\rm NS}/R_{\rm lc}}=\sqrt{R_{\rm NS} \Omega/c}=\sqrt{2\pi R_{\rm NS}/(c P)}$ from the star's pole. The $B$-field lines that originate in the region between $\theta=0$ and $\theta=\theta_{\rm pc}$ (referred to as { the} \emph{magnetic polar caps}) cross the light cylinder and are called ``open'' field lines. Charged particles leave the star moving along the open field lines and { escape}  from the magnetosphere passing through the light cylinder.

At large distances from the light cylinder the magnetic field lines becomes radial. { Thus, } the magnetic field geometry is dominated by the toroidal component which decreases with the inverse of the distance. For typical pulsar magnetospheres it is expected to be related { to}  the poloidal component of the field at the surface, $B_s$, as \citep[see][for details]{1969ApJ...157..869G}
\begin{equation}\label{eq:BtBs}
B_t\sim \left(\frac{2\pi R_{\rm NS}}{c P}\right)^2 \left(\frac{R_{\rm NS}}{r}\right)B_s,
\end{equation}
up to a factor of order unity. Thus, as the SN remnant expands it finds a magnetized medium with a different value of the $B$-field.
We adopt a magnetic field of the form
\begin{equation}\label{eq:Bt}
B(t)= B_0 \left(\frac{R_0}{r}\right)^{-m},
\end{equation}
with $1\leq m \leq2$. We then seek the value of $m$ which fits best the data (see Secs.~\ref{sec:5}--\ref{sec:7}).

According to the previous agreement we have found between our model and GRB data \citep[see e.g.][]{2016ApJ...833..107B,2018ApJ...852...53R}, we shall adopt values for $R_0$ and the expansion velocity $\dot{R}$ (see below Secs.~\ref{sec:5}--\ref{sec:7}) and leave the parameter $B_0$ to be set by the fit of the afterglow data. We then compare and contrast the results with that expected from the NS theory.

%%%%%%%%%%%%%%%%%%%%%%%%%%%%%%%%%%%%%%%%%%%%%%%%%%%%%
%%%%%%%%%%%%%%%%%%%%%%%%%%%%%%%%%%%%%%%%%%%%%%%%%%%%%
\section{Model for the Optical and X-ray Spectrum of the Afterglow}\label{sec:5}
%%%%%%%%%%%%%%%%%%%%%%%%%%%%%%%%%%%%%%%%%%%%%%%%%%%%%
%%%%%%%%%%%%%%%%%%%%%%%%%%%%%%%%%%%%%%%%%%%%%%%%%%%%%

The origin of the observed afterglow emission is interpreted { here} as due to the synchrotron emission of electrons accelerated in an expanding magnetic HN ejecta.\footnote{We note that synchrotron emission of electrons in fast cooling regime has been previously applied in GRBs but to explain the prompt emission \citep[see e.g.][]{2014NatPh..10..351U}.}  A fraction of the kinetic energy of the ejecta is converted, through a shockwave, to accelerated particles (electrons) above GeV and TeV energies --- enough to emit photons up to the X-ray band by synchrotron emission. Depending on the shock speed, number density, magnetic field, etc., different initial energy spectra of particles can be formed. In the most common cases, the accelerated particle distribution function can be described by a power law in the form of
\begin{equation}
Q(\gamma,t)=Q_0(t)\gamma^{-p}\theta(\gamma_\mathrm{max}-\gamma)\theta(\gamma-\gamma_\mathrm{min}) \, ,
\label{PL}
\end{equation}
where $\gamma = E/{m c^2}$ is the electron Lorentz factor, $\gamma_\mathrm{min}$ and $\gamma_\mathrm{max}$ are the minimum and maximum Lorenz factors, respectively. $Q_0(t)$ is the number of injected particles per second per energy, originating from the remnant impacted by the $e^+e^-$ pair plasma of the GRB.

After the electrons are injected with { a}  spectrum given by Eq.~(\ref{PL}), the evolution of the particle distribution at a given time can be determined from the solution of the kinetic { equation  of}  the electrons taking into account the particle energy losses \citep{1962SvA.....6..317K}
\begin{equation}
\frac{\partial N(\gamma,t)}{\partial t} = \frac{\partial}{\partial \gamma} (\dot{\gamma}(\gamma,t) \, N(\gamma,t))-\frac{N(\gamma,t)}{\tau}+Q(\gamma,t) \, ,
\label{Neq}
\end{equation}
where $\tau$ is the characteristic escape time and $\dot{\gamma}(\gamma,t)$ is the cooling rate. In the present case { the escape time for electrons} is much longer than the characteristic { cooling time scale} (fast cooling regime). The term $\dot{\gamma}(\gamma,t)$ includes various electron energy loss processes, such as synchrotron and inverse-Compton cooling as well as adiabatic losses due to the expansion of the emitting region. For the magnetic field considered here, the dominant cooling process for higher energy electrons is synchrotron emission (the electron cooling timescale due to inverse-Compton scattering is significantly longer) while adiabatic cooling can dominate for the low energy electrons at later phases. By introducing the expansion velocity of the remnant $\dot{R}(t)$ and its radius $R(t)$, the energy loss rate of electrons can be written as
\begin{equation}
\dot{\gamma}(\gamma,t)=\frac{\dot{R}(t)}{R(t)}\gamma+\frac{4}{3} \frac{\sigma_\mathrm{T}}{m_\mathrm{e}c}\frac{B(t)^2}{8\pi}\gamma^2 \, ,
\end{equation}
where $\sigma_\mathrm{T}$ is the Thomson cross section and $B(t)$ is the magnetic field strength. From  the early X-ray data we find that the initial expansion velocity of GRB 130427A at times $\sim 10^2$~s is $0.8c$ \citep{2015ApJ...798...10R}, which then decelerates to $0.1c$ at $10^6$~s, as inferred from the SN optical data \citep{2013ApJ...776...98X}. 

{ Supernova or hypernova remnants like the one considered here generally evolve through three stages \citep[see][]{1997ApJ...490..619S}. These are  the free expansion phase, the Sedov phase, and the radiative cooling phase.  The free expansion phase roughly ends when the total mass of gas swept up by the shock equals the initial supernova ejecta mass. During this phase, the shock velocity remains nearly constant at its initial velocity $v_0$ and the outer radius $R$ of the ejecta evolves linearly in time after the explosion. This phase ends \citep{1997ApJ...490..619S} when 
\begin{equation}
t \approx  50{\rm~ yr} \times \biggl[ \biggl (\frac{M_{\rm ej}}{5 M_\odot}\biggr) \times  \biggl(\frac{1~{\rm cm^{-3}}}{ n_{\rm ISM}}\biggr) \times  \biggl(\frac{v_0}{0.1~c}\biggr)^3\biggr]^{1/3}~~,
\end{equation}
where $M_{\rm ej}$ is the HN ejected mass and  $n_{\rm ISM}$ is the hydrogen density in the local interstellar medium.  For a mildly relativistic ejecta ($v/c \sim  0.9$, $\Gamma \sim 3$) in a typical ISM of $n_{\rm ISM}  \approx  1$ cm$^-3$ this phase lasts for 450 years.  Even if the ISM is 1000 times more dense due to past mass loss of the progenitor star, this phase still lasts for 45 years.  Since we only consider times much less than a year (out to $10^7$ sec) we are completely justified in treating the expansion as  a ``ballistic'' constant velocity rather than a Sedov expansion. 

Nevertheless, we allow for an initial  linearly decelerating eject as observed in the thermal component (cf. Sec. 3)) until $10^6$~s.  After which it is allowed to expand with a constant velocity of $0.1c$. Thus,  the expansion velocity of { the }ejecta is written as}
	\begin{eqnarray}\label{eqn:expansion}
	\dot{R}(t) & = & \begin{cases} 
	v_0 - a_0 \, t & t\leq 10^6 \mathrm{s} \\
	v_f & t > 10^6 \mathrm{s} 
	\end{cases} \, , \\
	R(t) & = & \begin{cases} 
	v_0 \, t - a_0 \, t^2/2 & t\leq 10^6 \mathrm{s} \\
	1.05 \times 10^{16}\,\mathrm{cm}+ v_f\, t & t > 10^6 \mathrm{s} 
	\end{cases} \, ,
	\end{eqnarray}
where $v_0 = 2.4 \times 10^{10}$~cm~s$^{-1}$, $a_0=2.1 \times 10^{4}$~cm~s$^{-2}$, and $v_f = 3 \times 10^{9}$~cm~s$^{-1}$. 

Due to the above decelerating expansion of the emitting region, the magnetic field decreases. Therefore we adopt a magnetic field that scales as $B(t)=B_0 \left(\frac{R(t)}{R_0}\right)^{-m}$
 with $1\leq m \leq2$. We shall show below (see Sec.~\ref{sec:7}) that the data  { are} best fit with $m=1$. This corresponds to conservation of magnetic flux for the longitudinal component.

The initial injection rate of particles, $Q_0(t)$, depends on the energy budget of ejecta and on the efficiency of converting from kinetic to non-thermal energy. This can be defined as
\begin{equation}
L(t)=Q_0(t) m_e c^2\int_{\gamma_\mathrm{min}}^{\gamma_\mathrm{max}}\gamma^{1-p} d\gamma \, ,
\end{equation}
where it is assumed that $L(t)$ varies in time, based on the recent analyses of BdHNe which show that the X-ray light curve of GRB 130724A decays in time following a power-law of index $\sim -1.3$ (\citealp{2015ApJ...798...10R}; see Fig.~\ref{fig:slope}). In our interpretation, the emission in the optical and X-ray bands is produced from synchrotron emission of electrons: if one assumes the electrons are constantly injected ($L(t)=L$), this will produce  { a } constant synchrotron flux. Thus, we assume that the  luminosity of  { the electrons changes} from an initial value $L_0$ as follows: 
\begin{equation}\label{eq:Lt}
L(t)=L_0 \times \left(1+\frac{t}{\tau_0}\right)^{-k},
\end{equation}
{ where the $L_0$ and $k$ are fixed by the observed afterglow light curve (see Eq.~\ref{eq:Lsyn})  (see details below in Secs.~\ref{sec:6} and \ref{sec:7})}.

The kinetic equation given in Eq.~(\ref{Neq}) has been solved numerically. The discretized electron continuity equation (\ref{Neq}) is re-written in the form of a tridiagonal matrix which is solved using the implementation of the ``tridiag'' routine in \citet{1992nrca.book.....P}. We have carefully tested our code by comparing the numerical results with the analytic solutions given in \citet{1962SvA.....6..317K}.

The synchrotron luminosity temporal evolution is calculated using $N(\gamma,t)$ with
\begin{equation}\label{eq:Lsyn}
L_{syn}(\nu,t)=\int_{1}^{\gamma_\mathrm{max}}{N(\gamma,t)P_{syn}(\nu,\gamma,B(t))d\gamma},
\end{equation}
where $P_{syn}(\nu,\gamma,B(t))$ is the synchrotron spectra for a single electron which is calculated using the parameterization of the emissivity function of synchrotron radiation presented in \citet{2010PhRvD..82d3002A}.

%%%%%%%%%%%%%%%%%%%%%%%%%%%%%%%%%%%%%%%%%%%%%%%%%%%%%%%%%%%
%%%%%%%%%%%%%%%%%%%%%%%%%%%%%%%%%%%%%%%%%%%%%%%%%%%%%%%%%%%
\section{Initial Conditions for GRB 130724A}\label{sec:6}
%%%%%%%%%%%%%%%%%%%%%%%%%%%%%%%%%%%%%%%%%%%%%%%%%%%%%%%%%%%
%%%%%%%%%%%%%%%%%%%%%%%%%%%%%%%%%%%%%%%%%%%%%%%%%%%%%%%%%%%

In \citet{2018ApJ...852...53R} an analysis was completed for seven subclasses of GRBs including 345 identified BdHNe candidates, one of which is  GRB 130724A that was seen in the {\it Swift}-XRT data and analyzed in detail in \citet{2015ApJ...798...10R}. From the host-galaxy identification it is known that this burst occurred at a redshift $z = 0.334$. After transforming to the cosmological rest-frame of the burst and properly correcting for effects of the cosmological redshift and Lorentz time dilation, one can infer a time duration $t_{90} = 162.8$~s for 90\% of the GRB emission. The isotropic energy emission in the range of $1$--10$^4$~keV in the cosmological rest-frame of the burst is also deduced to be $E_{iso} = (9.3 \pm 1.3) \times 10^{53}$~erg and the total emission in the power-law afterglow can be inferred \citep{2015ApJ...798...10R}.  { This fixes $L_0$ in Eq.~(\ref{eq:Lt}).}

\begin{figure}
	\centering
	\includegraphics[width=1.1\hsize,clip]{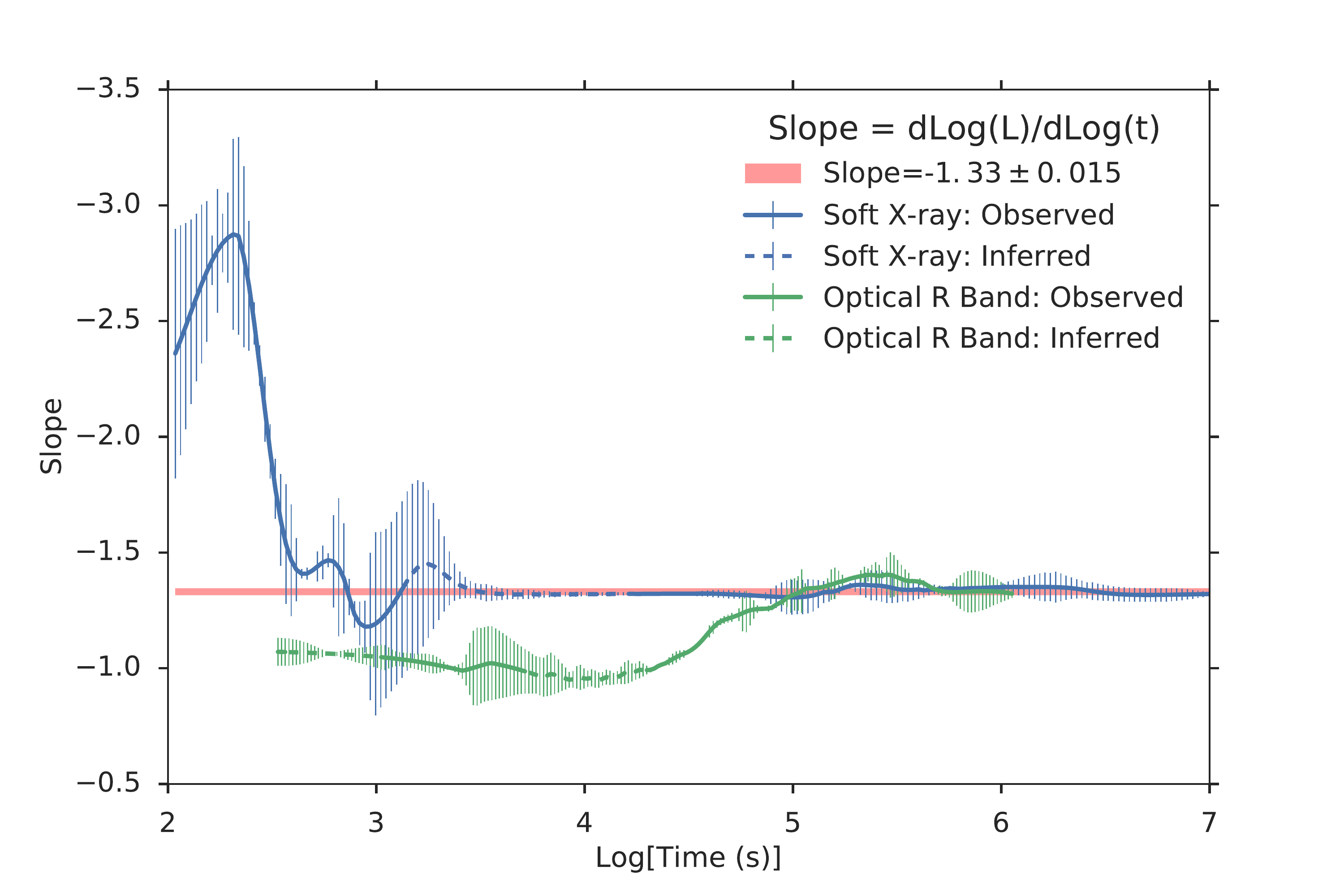}
	\caption{The slope of the afterglow light-curve of BdHN 130427A, defined by the logarithmic time derivative of the luminosity: slope = $d \log_{10}(L)/ d \log_{10}(t)$. This slope is obtained by fitting the luminosity light-curve in the cosmological rest-frame, using a machine learning, locally weighted regression (LWR) algorithm. For the corresponding technical details and codes we refer the reader to: \url{https://github.com/YWangScience/AstroNeuron}. The green line is the slope of the soft X-ray emission, in the $0.3$--$10$~keV range, and the blue line corresponds to the optical R-band, centered at $658$~nm.}
	\label{fig:slope}
\end{figure}

Fig.~\ref{fig:slope} shows the slope of the light-curve, defined by the logarithmic time derivative of the luminosity: slope = $d \log_{10}(L)/ d \log_{10}(t)$. This slope is obtained by fitting the luminosity light-curve in the cosmological rest-frame, using a machine learning, locally weighted regression (LWR) algorithm. We have made publicly available the corresponding technical details and codes to perform this calculation at: \url{https://github.com/YWangScience/AstroNeuron}. The green line is the slope of the soft X-ray  { emission}, in the $0.3$--$10$~keV range, and the blue line corresponds to the optical R-band, centered at $658$~nm. The solid line covers the time when the data are well observed, while the dashed line, corresponds to an epoch in which observational data are missing. The rapid change of the slope implies variations of the energy injection, different emission mechanisms or different emission phases. The slope of the soft X-ray  { emission  varies dramatically} at early times when various complicated GRB components (prompt emission, gamma-ray flare, X-ray flare) are occurring. Hence, we do not attempt to explain this early part with the synchrotron emission model defined above. We only consider times later than $10^3$~s. Also we note that, at times later than $10^5$~s, the slopes of the X-ray and R bands reach a common value of $-1.33$, indicated as a red line.

Furthermore, we are not interested in explaining the GeV emission observed in most of BdHNe (when LAT data are available) with the synchrotron radiation model proposed here. Such emission has been explained in \citet{2015ApJ...798...10R} as originating from the further accretion of
matter onto the newly-formed BH. This explanation is further reinforced by the fact that a similar GeV emission, following the same power-law decay with time, is also observed in the authentic short GRBs (S-GRBs; short bursts with $E_{iso} \gtrsim 10^{52}$~erg; see \citealp{2016ApJ...832..136R}) which are expected to be produced in NS-NS mergers leading to BH formation (\citealp{2016ApJ...832..136R}; Aimuratov et al., in preparation).

Regarding the model parameters, the initial velocity of the expanding ejecta is expected to be $v_0=2.4 \times 10^{10}$~cm~s$^{-1}$ \citep{2015ApJ...798...10R} from the thermal black body emission. Similarly, the radius at the beginning of the X-ray afterglow should be $R_0 \approx 2.4 \times 10^{12}$~cm. This corresponds to an expansion timescale { of $t_0 = \tau_0 = 100$~s. }These values are consistent with our previous theoretical simulations of BdHNe \citep{2016ApJ...833..107B}. For our simulation of this burst we include all expected energy losses (synchrotron and adiabatic energy losses).  However, the escape timescale was assumed to be large so that its effect could be neglected.
 
%%%%%%%%%%%%%%%%%%%%%%%%%%%%%%%%%%%%%%%%%%%%%%%%
%%%%%%%%%%%%%%%%%%%%%%%%%%%%%%%%%%%%%%%%%%%%%%%%
\section{Results}\label{sec:7}
%%%%%%%%%%%%%%%%%%%%%%%%%%%%%%%%%%%%%%%%%%%%%%%%
%%%%%%%%%%%%%%%%%%%%%%%%%%%%%%%%%%%%%%%%%%%%%%%%

\begin{figure}
	\centering
	\includegraphics[width=\hsize,clip]{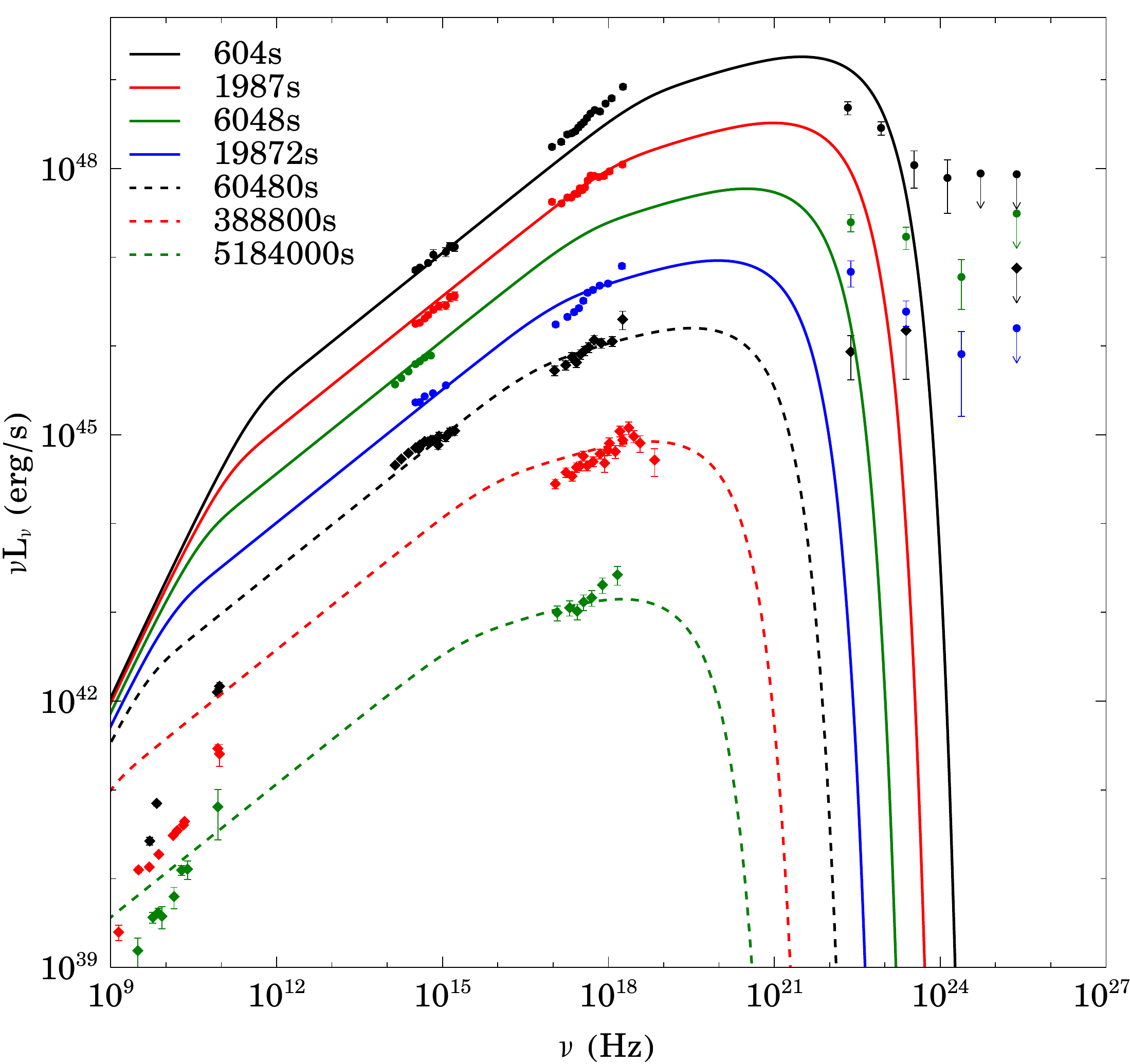}
	\caption{Model evolution (lines) of synchrotron spectral luminosity at various times compared with measurements (points with error bars)  in various spectral bands for GRB 130724A.}
	\label{fig:movinglimitsmodelb05e2min1e3max5e5}
\end{figure}

Our modeling  { of} the broadband spectral energy distribution (SED) of GRB 130724A for different periods is shown in Fig.~\ref{fig:movinglimitsmodelb05e2min1e3max5e5}. The corresponding parameters are given in Table~\ref{tab:parameters}. { However, as noted above the  8  parameters in Table~\ref{tab:parameters} are not all ``free'' and independent. For example, $R_0$ and $t_0 = \tau_0$ are fixed by the observed thermal component.  Also, $\gamma_{\rm min}$  and $\gamma_{\rm max}$ are fixed once $B$ is given. $L_0$ is fixed by a normalization of the observed source luminosity.  The synchrotron index $p$ is not varied, but kept fixed at 1.5 as typical of synchrotron emission.  The parameter $ k$ is fixed by the slope of the late time X-ray afterglow.   Hence, the only ``free parameter'' is $B_0$.  This parameter then provides an excellent fit to the observed spectra and light curves over a broad range of wavelengths and time scales for a single plausible value.}

The radio emission is due to low-energy electrons that accumulate for longer periods. That is why the radio data are not included in the model. Only the optical and X-ray emissions are interpreted as due to { the }synchrotron emission of electrons. Such emission, for instance at 604~s, is produced in a region with a radius of $1.4\times10^{14}$~cm and a magnetic field of $B=8.3 \times 10^4$~G. For this field strength synchrotron self-absorption can be significant as estimated following \citet{1979rpa..book.....R}. At the initial phases, when the system is compact and the magnetic field is large, synchrotron-self absorption can be neglected for the photons with frequencies above $10^{14}$~Hz. { Otherwise,} it is important. Thus, it is effective in reducing the radio flux predicted by the model, but not the optical and X-ray emission.

The optical and X-ray data can be well fit by a single power-law injection of electrons with $Q\propto \gamma^{-1.5}$ and with initial minimum and maximum energies of $\gamma_{\rm min}=4\times10^3$ ($E_{\rm min}=2.0$~GeV) and $\gamma_{\rm max}=5\times10^5$ ($E_{\rm max}=255.5$~GeV), respectively. Due to the fast synchrotron cooling, the electrons are cooled rapidly forming a spectrum of $N(\gamma,t)\sim\gamma^{-2}$ for $\gamma \leq \gamma_{\rm min}$ and $N(\gamma,t)\sim\gamma^{-2.5}$ for $\gamma \geq \gamma_{\rm min}$. The slope of the synchrotron emission ($\nu F_{\nu}\propto \nu^{1-s}$) below the frequency defined by $\gamma_{\rm min}$ (e.g., $h\:\nu_{\rm min}\simeq3\:e\:h\:B(t)\:\gamma_{\rm min}^2/4\:\pi\:m_{e}\:c$) { is $s=(2-1)/{2}=0.5$.} This explains well both the optical and X-ray data. 

\begin{figure}
	\centering
	\includegraphics[width=\hsize,clip]{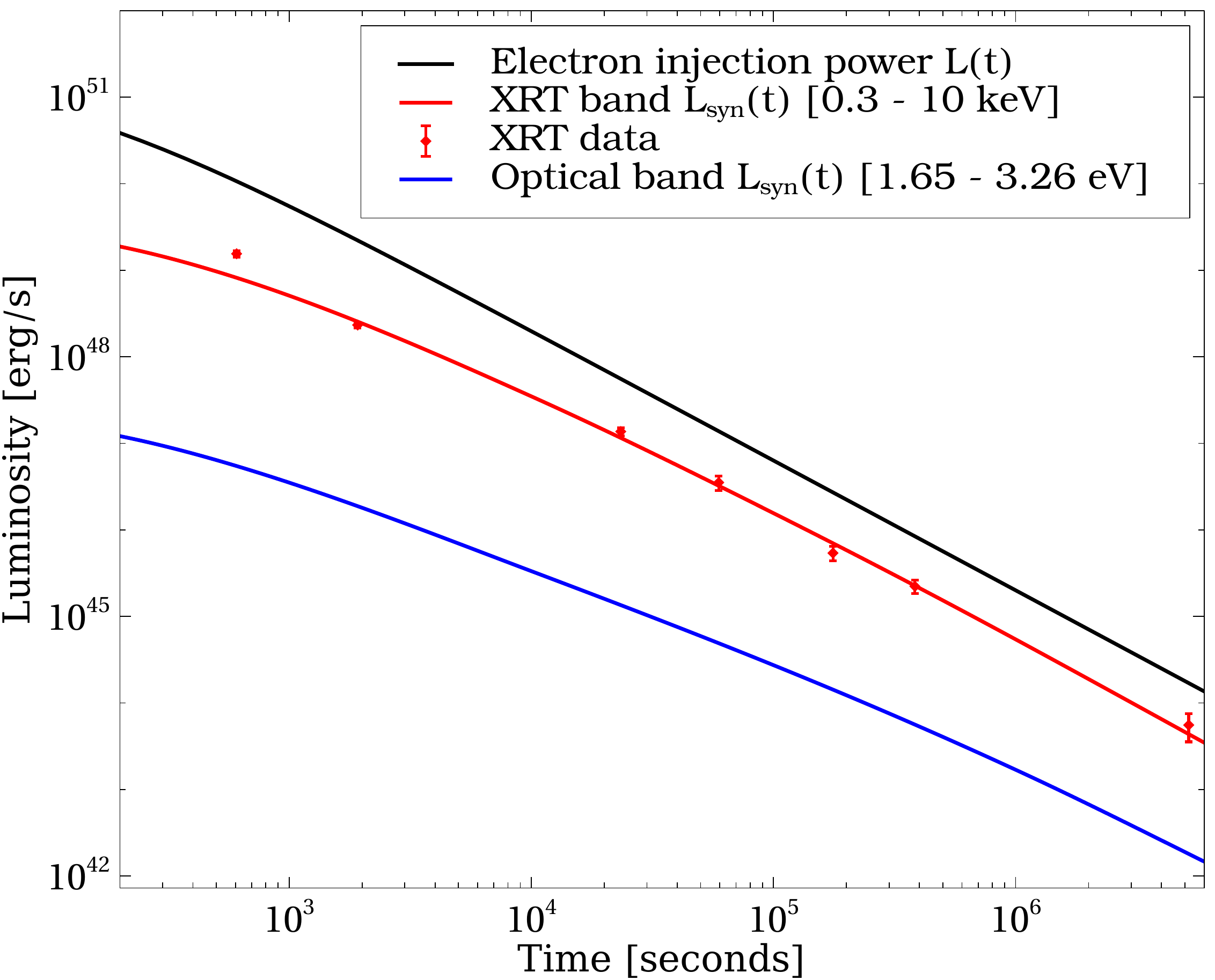}
	\caption{{X-ray light-curve of GRB 130427A (points with error bars) together with the optical and X-ray} theoretical synchrotron light-curve (lines) from Eq.~(\ref{eq:Lsyn}). We also show the electron injection power $L(t)$ given by Eq.~(\ref{eq:Lt}).}
	\label{fig:Lx}
\end{figure}

For frequencies above $\nu_{\rm min}$, the slope is $\nu F_{\nu}\propto \nu^{0.25}$ which continues up to $h\:\nu_{\rm max}\simeq3\:e\:h\:B(t)\:\gamma_{\rm max}^2/(4 \pi m_e c)$. Since $\nu_{\rm min}$ and $\nu_{\rm max}$ depend on the magnetic field, they decrease with time, e.g. at $t=5.2\times10^6$~s, $\nu_{\rm min}\simeq6.5\times10^{14}$~Hz and $\nu_{\rm max}\simeq1.0\times10^{19}$~Hz. Due to the changes in the initial particle injection rate and magnetic field, the synchrotron luminosity also decreases. This is evident from Fig.~\ref{fig:Lx}, where the observed optical and X-ray light-curves of GRB 130427A are compared with the theoretical synchrotron emission light-curve obtained from Eq.~(\ref{eq:Lsyn}). In this figure we also show the electron injection power $L(t)$ given by Eq.~(\ref{eq:Lt}). Here, it can be seen how the synchrotron luminosity fits the observed decay of the afterglow luminosity with the correct power-law index $~-1.3$ (see also Fig.~\ref{fig:slope}).

\begin{deluxetable}{c|c} 
\tablecaption{Parameters used for the simulation of GRB 130724A.\label{tab:parameters}}
\tablehead{\colhead{Parameter} & \colhead{Value}} 
\startdata
$B_0$ & $5.0 (\pm 1) \times 10^5 \; \mathrm{G}$ \\
$R_0$ & $2.4\times 10^{12}\; \mathrm{cm}$ \\
$L_{0}$ & $2.0 \times10^{51} \; \mathrm{erg/s}$  \\
$k$ & $1.58$ \\
$\tau_0$ & $1.0 \times 10^2 \; \mathrm{s}$\\
$p$ & $1.5$\\
$\gamma_\mathrm{min}$ & $4.0 \times 10^3$\\
$\gamma_\mathrm{max}$ & $5.0 \times 10^5$\\
\enddata
\end{deluxetable}
     
The SN ejecta is expected to become transparent to the $\nu$NS radiation at around $10^5$~s. Thus, we now discuss the pulsar emission that might power the late ($t\gg 10^5$~s) X-ray afterglow light-curve. 

The late X-ray afterglow also shows a power-law decay of index $\sim -1.3$ which, as we show below, if powered by the pulsar implies the presence of a quadrupole magnetic field in addition to the traditional dipole one. 

Thus, we adopt a dipole+quadrupole magnetic field model \citep[see][for details]{2015MNRAS.450..714P}. The luminosity from a pure dipole ($l=1$) is
\begin{equation}
	L_{dip} = \frac{2}{3 c^3} \Omega^4 B_{dip}^2 R_{\rm NS}^6 \sin^2\chi_1,
\end{equation}
where $\chi_1$ = 0 degrees gives the axisymmetric mode $m = 0$ alone whereas $\chi_1$ = 90 degrees gives the $m = 1$ mode alone. The braking index, following the traditional definition $n \equiv \Omega \ddot{\Omega}/\dot{\Omega}^2$, is in this case $n = 3$. 

On the other hand, the luminosity from a pure quadrupole field ($l=2$) is
\begin{equation}
	L_{quad} = \frac{32}{135 c^5} \Omega^6 B_{quad}^2 R_{\rm NS}^8 \sin^2\chi_1(\cos^2\chi_2+10\sin^2\chi_2),
\end{equation}
where the different modes are easily separated by taking $\chi_1$ = 0 and any value of $\chi_2$ for $m = 0$, ($\chi_1$, $\chi_2$) = (90, 0) degrees for $m = 1$ and ($\chi_1$, $\chi_2$) = (90, 90) degrees for $m = 2$. The braking index in this case is $n=5$.

Thus, the quadrupole to dipole luminosity ratio is:
\begin{equation}
	R^{quad}_{dip} = \eta^2 \frac{16}{45} \frac{R_{\rm NS}^2 \Omega^2}{c^2},
\label{eq:ratio}
\end{equation}
where
\begin{equation}
    \eta^2 = (\cos^2\chi_2+10\sin^2\chi_2) \frac{B_{quad}^2}{B_{dip}^2}.
\label{eq:eta}
\end{equation}

It can be seen that $\eta = B_{quad}/B_{dip}$ for the $m=1$ mode, and $\eta = 3.16 \times B_{quad}/B_{dip}$ for the $m=2$ mode. For a $1$~ms period $\nu$NS, if $B_{quad} = B_{dip}$, the quadrupole emission is about $\sim 10\%$ of the dipole emission, if $B_{quad} = 100 \times B_{dip}$, the quadrupole emission increases to $1000$ times the dipole emission; and for a $100$~ms pulsar, the quadrupole emission is negligible when $B_{quad} = B_{dip}$, or only  $\sim 10\%$ of the dipole emission even when $B_{quad} = 100 \times B_{dip}$. From this result one infers that the quadrupole emission dominates in the early fast rotation phase, then the $\nu$NS spins down and the quadrupole emission drops faster than the dipole emission and, after tens of years, the dipole emission becomes the { dominant} component.

The evolution of the $\nu$NS rotation and  luminosity are given by 
\begin{eqnarray}
	\frac{dE}{dt} &=& -I \Omega  \dot{\Omega } =  - (L_{dip} + L_{quad}) \nonumber \\
    &=& - \frac{2}{3 c^3} \Omega^4 B_{dip}^2 R_{\rm NS}^6 \sin^2\chi_1 \left(1+\eta^2 \frac{16}{45} \frac{R_{\rm NS}^2 \Omega^2}{c^2}\right),
\end{eqnarray}
where $I$ is the moment of inertia. The solution is
\begin{equation}
	t = f(\Omega) - f(\Omega_0)
\label{eq:tOmega}
\end{equation}
where
\begin{equation}
	f(\Omega) = \frac{3 I c \{\frac{16}{45} \eta^2  R_{\rm NS}^2 \Omega ^2 [2 \ln \Omega-\ln (c^2+ \frac{16}{45}\eta^2  R_{\rm NS}^2 \Omega ^2)]+c^2\}}{4 B_{dip}^2 \sin^2\chi_1 R_{\rm NS}^6 \Omega ^2}
\end{equation}
and
\begin{equation}
	f(\Omega_0) = \frac{3 I c \{\frac{16}{45} \eta^2  R_{\rm NS}^2 \Omega_0 ^2 [2 \ln \Omega_0-\ln (c^2+ \frac{16}{45}\eta^2  R_{\rm NS}^2 \Omega_0 ^2)]+c^2\}}{4 B_{dip}^2 \sin^2\chi_1 R_{\rm NS}^6 \Omega_0 ^2}
\end{equation}

The first and the second derivative of the angular velocity are
\begin{equation}
	\dot{\Omega } = -\frac{2 B_{dip}^2 \sin^2\chi_1 R_{\rm NS}^6 \Omega^3}{3 I c^3} (1+\eta^2 \frac{16}{45 c^2} R_{\rm NS}^2 \Omega^2)
\label{eq:dotOmega}
\end{equation}
\begin{equation}
    \ddot{\Omega } = - \frac{2 B_{dip}^2 \sin^2\chi_1 R_{\rm NS}^6 \Omega^2 \dot{\Omega}}{I c^3}(1+\eta^2 \frac{16}{27 c^2} R_{\rm NS}^2 \Omega^2  )
\end{equation}
Therefore the braking index is
 \begin{equation}
     n = \frac{\Omega \ddot{\Omega }}{\dot{\Omega }^2} = \frac{135c^2+80\eta^2 R_{\rm NS}^2 \Omega^2}{45c^2+16\eta^2 R_{\rm NS}^2 \Omega^2}
 \end{equation}
that in the present case ranges from $3$ to $5$. From Eqs.~(\ref{eq:tOmega}--\ref{eq:dotOmega}) we can compute the evolution of total pulsar luminosity as
\begin{equation}
	L_{tot}(t) = I \Omega \dot{\Omega }.
\end{equation}

\begin{figure}
\centering
\includegraphics[width=1.1\hsize,clip]{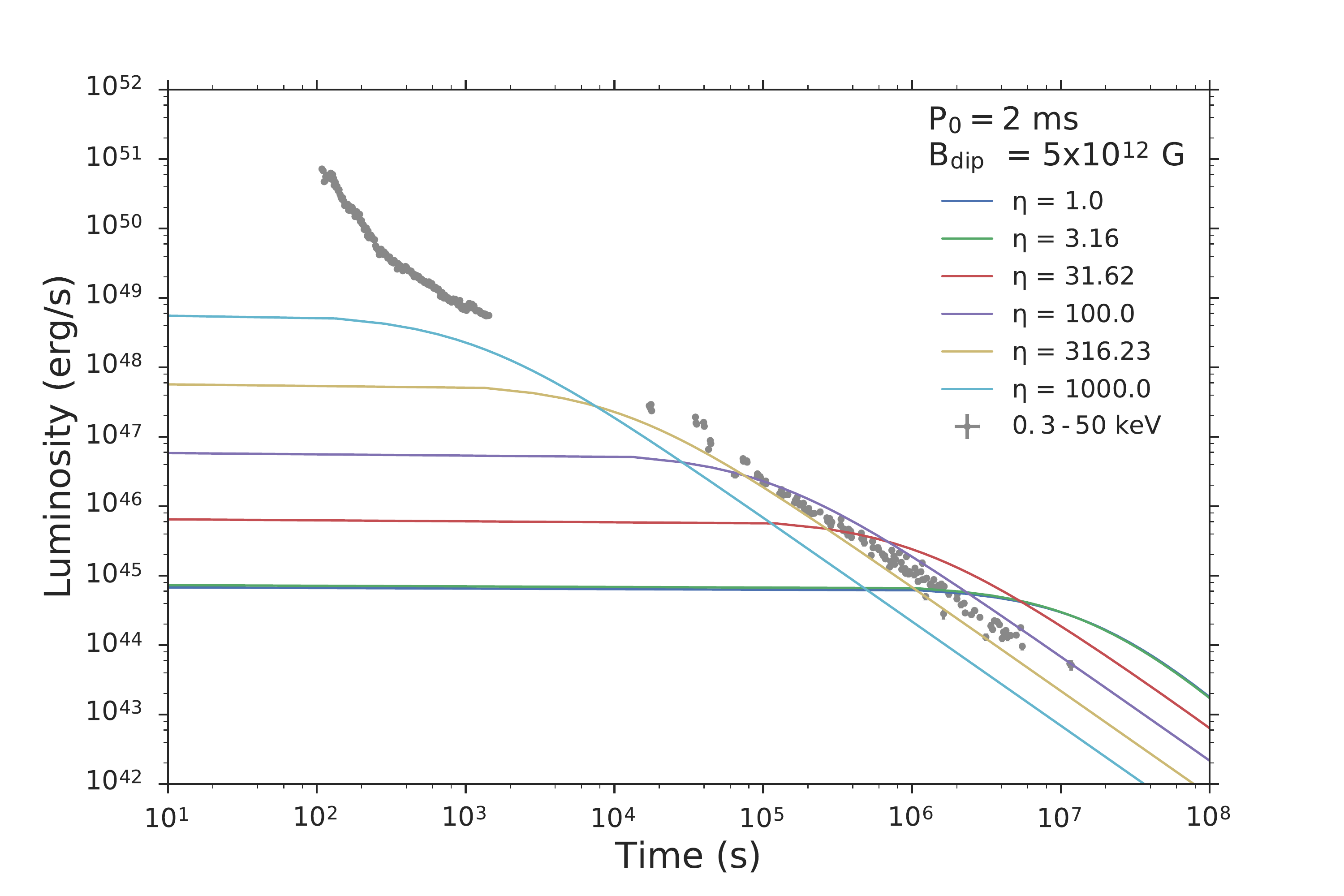}
\caption{The observed luminosity of GRB 130427A in the $0.3$--$50$~keV band (grey points), and the theoretical luminosity from a pulsar for selected quadrupole to dipole magnetic field ratio and quadrupole angles in color lines. Other parameters of the pulsar are fixed: initial spin period $P_0 = 2$~ms, dipole magnetic field $B_{dip} =  5 \times 10^{12}$~G, inclination angle $\chi_1 = \pi/2$, mass $M = 1.5~M_\odot$, radius $R_{\rm NS} = 1.5 \times 10^6$~cm.}\label{fig:Lpulsar}
\end{figure}

Figure~\ref{fig:Lpulsar} shows the luminosity obtained from the above model for a $1.5~M_\odot$ pulsar with a radius of $1.5\times10^6$~cm, $B_{dip} = 5\times10^{12}$~G, an initial rotation period $P_0 = 2$~ms, and for selected values of the parameter $\eta$. This figure shows that the theoretical luminosity of { the }pulsar is close to the soft X-ray luminosity observed in GRB 130427A when $\eta$ is around $100$. This means, if choosing the harmonic mode $m=2$, the quadrupole magnetic field is about $30$ times stronger than the dipole magnetic field. The luminosity of the pulsar before $10^6$~s is mainly powered by the quadrupole emission, which is tens of times higher than the dipole emission. At about $10$ years the dipole emission starts to surpass the quadrupole emission and continues to dominate thereafter.

It is important to check the self-consistency of the estimated $\nu$NS parameters obtained first from the early afterglow via synchrotron emission and then from the late X-ray afterglow via the pulsar luminosity. We can obtain from Eqs.~(\ref{eq:Bt}) and (\ref{eq:BtBs}), via the values of $B_0$ and $R_0$ from Table~\ref{tab:parameters} and for $P_0 = 2$~ms, an estimate of the dipole field at the $\nu$NS surface from the synchrotron emission powering the early X-ray afterglow, $B_s \approx 6.7\times 10^{12}$~G. This value is to be compared with the one we have obtained from the pulsar luminosity powering the late afterglow, $B_{dip} = 5\times 10^{12}$~G. The self-consistency of the two estimates is remarkable. In addition, the initial rotation period $P_0 = 2$~ms for the $\nu$NS is consistent with our estimate in Sec.~\ref{sec:4} based upon angular momentum conservation during the gravitational collapse of the iron core leading to the $\nu$NS. It can also be checked from Fig.~\ref{fig:ErotvsM} that $P_0$ is longer than the minimum period of a $1.5~M_\odot$ NS, which guarantees the gravitational and rotational stability of the $\nu$NS.

%%%%%%%%%%%%%%%%%%%%%%%%%%%%%%%%%%%%%%%%%%%%%%%%
%%%%%%%%%%%%%%%%%%%%%%%%%%%%%%%%%%%%%%%%%%%%%%%%
\section{Conclusions}\label{sec:8}
%%%%%%%%%%%%%%%%%%%%%%%%%%%%%%%%%%%%%%%%%%%%%%%%
%%%%%%%%%%%%%%%%%%%%%%%%%%%%%%%%%%%%%%%%%%%%%%%%

We have constructed a model for a broad frequency range of the observed spectrum in the afterglow of BdHNe. We have made a specific fit to the BdHN 130427A as a representative example. We find that the parameters of the fit are consistent with the BdHN interpretation for this class of GRBs.

We have shown that the optical and X-ray emission of the early ($ 10^2$~s$\lesssim t\lesssim 10^6$~s) afterglow is explained by the synchrotron emission { from} electrons expanding in the HN threading the magnetic field of the $\nu$NS. At later times the HN becomes transparent and the electromagnetic radiation from the $\nu$NS dominates the X-ray emission. We have inferred that the $\nu$NS possesses an initial rotation period of 2~ms and a dipole magnetic field of (5--7)$\times 10^{12}$~G. It is worth mentioning that we have derived the strength of the magnetic dipole independently by the synchrotron emission model at early times ($t\lesssim 10^6$~s) and by the magnetic braking model powering the late ($t\gtrsim 10^6$~s) X-ray afterglow and show that they are in full agreement.

In this paper we proposed a direct connection between the afterglow of a BdHN and the physics of a newly born fast-rotating NS. This establishes a new self-enhancing understanding both of GRBs and young SNe which could be of fundamental relevance for the understanding of ultra-energetic cosmic rays and neutrinos as well as new ultra high energy phenomena.

It appears to be now essential to extend our comprehension in three different directions: 1) understanding of the latest phase of the afterglow; 2) the possible connection with historical supernovae;  as well as 3) to extend observations from space of the GRB afterglow in the GeV and TeV energy bands. These last observations are clearly additional to the current observations of GRBs and GRB GeV radiation, originating from a Kerr-Newman BH and totally unrelated to the { astrophysics} of afterglows.   

One of the major verifications of our model can come from observing, in still active afterglows of historical GRBs, the pulsar-like emission from the $\nu$NS we here predict, and the possible direct relation of the Crab Nebula to a BdHN is now open to further examination.

\acknowledgments
We acknowledge the continuous support of the MAECI. This work made use of data supplied by the UK Swift Science Data Center at the University of Leicester. M.K., is supported by the Erasmus Mundus Joint Doctorate Program Grant N.2014--0707 from EACEA of the European Commission. J.A.R. acknowledges the partial support of the project N. 3101/GF4 IPC-11, and the target program F.0679  0073-6/PTsF of the Ministry of Education and Science of the Republic of Kazakhstan.  Work at the University of Notre Dame (G.J.M.) is supported by the U.S. Department of Energy under Nuclear Theory Grant DE-FG02-95-ER40934. N.S. acknowledges the support of the RA MES State Committee of Science, in the frames of the research project No 15T-1C375.

\end{document}